\documentclass[aps,prl,twocolumn]{revtex4}
\usepackage{graphicx}
\usepackage{amsmath}
\begin{document}

\title{The loop-quantum-gravity vertex-amplitude\\[.5mm]}
\author{Jonathan Engle, Roberto Pereira, Carlo Rovelli}
\affiliation{Centre de Physique Th\'eorique de Luminy, Case 907, F-13288 Marseille, EU}
\date{\today}

\begin{abstract}\noindent
Spinfoam theories are hoped to provide the dynamics
of non-perturbative loop quantum gravity. But a number of  their
features remain elusive. The best studied one ---the euclidean
Barrett-Crane model---  does not have the boundary state space
needed for this, and there are recent indications that, consequently, 
it may fail to yield the correct low-energy $n$-point functions.
These difficulties can be traced to the $SO(4)\!\to\! SU(2)$
gauge fixing and the way certain second class constraints are imposed, 
arguably incorrectly, \emph{strongly}.   We present an alternative model,
that can be derived as a  \emph{bona fide} quantization of a Regge
discretization of euclidean general relativity, and where
the constraints are imposed \emph{weakly}.  Its state space
is a natural subspace of the $SO(4)$ spin-network space and
matches the $SO(3)$ hamiltonian spin network space.
The model provides a long sought  $SO(4)$-covariant 
vertex amplitude for loop quantum gravity.

\end{abstract}

\maketitle

\noindent
The \emph{kinematics} of loop quantum gravity (LQG) provides a
well understood
background-independent language for a quantum theory of physical
space \cite{abhay,libro,lqg}.
The \emph{dynamics} of the theory is not understood as cleanly. Dynamics
is studied along two lines: hamiltonian (as in the Schr\"odinger
equation) \cite{thomas} or covariant  (as in Feynman's covariant quantum 
field theory).  We focus on the second. The key object that
defines the dynamics in this language is the vertex 
amplitude, like  the vertex  $e \gamma^\mu \!\sim\!\sim
\put(-2.5,2.5){\circle*{3}} \!\!\!\!<$ that defines the dynamics of perturbative 
QED. What is the vertex of LQG?

The spinfoam formalism \cite{spinfoams} is viewed as a
possible tool for answering this question.  It can be
derived in a remarkable number of distinct ways, which converge to the
definition of transition amplitudes as a Feynman sum over spinfoams. 
A spinfoam is a two-complex (union of 
faces, edges and vertices) colored with
quantum numbers (spins associated to faces and intertwiners
associated to edges); it can be loosely interpreted as a history
of a spin network (a colored graph). Its amplitude contains the
product of the amplitudes of each vertex, and thus the vertices play
a role similar to the vertices of Feynman's covariant QFT
\cite{spinfoam,mike1}.  This picture is nicely implemented in three
dimensions (3d) by the Ponzano-Regge model \cite{PoRe}, where the vertex
amplitude is given by the 6j Wigner symbol, which can be obtained as
a matrix element of the hamiltonian of 3d gravity  \cite{KaPe}.

Compelling and popular as it is, however, this picture has never been fully 
implemented in 4d.   The best studied model in the 4d euclidean context is the
Barrett-Crane (BC) model \cite{BC}.  This is simple and elegant, has remarkable 
finiteness properties \cite{finiteness}, and can be considered a modification of 
a topological $BF$ quantum field theory, by means of constraints ---called simplicity
constraints--- whose classical limit yields precisely the constraints that 
change $BF$ theory  into general relativity (GR). Furthermore, in the low-energy
limit some of its $n$-point functions appear to agree with those computed from 
perturbative quantum GR \cite{gravprop}.
However, the suspicion that something is wrong with the BC model has
long been agitated.  Its boundary state space is 
similar, but does not exactly match, that of loop quantum gravity; 
in particular the volume operator is ill-defined.  Worse, 
recent calculations appear to indicate that some $n$-point
functions fail to yield the correct low-energy limit \cite{matrice}.
All these problems are related to the way the \emph{intertwiner} quantum 
numbers (associated 
to the operators measuring angles between the faces bounding the elementary 
quanta of space) are treated: 
These quantum numbers are fully constrained in the BC model
by imposing  the simplicity constraints as strong operator
equations ($C_n \psi=0$). But these constraints are second class 
and imposing such constraints strongly may lead to the incorrect 
elimination of physical degrees of freedom \cite{dirac}. 

It is therefore natural to try to implement in 4d the general picture 
discussed above by correcting the BC model \cite{mike1,sergei}. 
In this letter we show that this is possible,  by properly 
imposing some of the constraints weakly 
($\langle\phi \,C_n\, \psi\rangle=0$), and 
that the resulting theory has remarkable features.   
First, its boundary quantum state space
matches \emph{exactly} the one of $SO(3)$ loop quantum gravity: no 
degrees of freedom are lost.
Second, as the degrees of freedom missing in BC are recovered, the 
vertex may yield the correct low-energy $n$-point functions.  Third,
the vertex can be seen as a vertex over $SO(3)$ spin networks or
$SO(4)$ spin networks, and is both $SO(3)$ and $SO(4)$ covariant.
Finally, the theory can be obtained as a bona fide quantization 
of a discretization of euclidean GR on a Regge triangulation.   Here we 
give  the definition of the theory, we illustrate its main aspects and we give
only a rapid sketch of its derivation from Regge GR. Details will be given 
elsewhere.

\vskip.3cm 
The model we discuss is defined by a standard spinfoam partition function
\begin{equation}
Z_{\rm GR}= \sum_{j_f, i_e}\ \prod_f (\dim {\scriptstyle\frac{j_f}{ 2}})^2\ 
\prod_v A(j_f,i_e)
\label{Z}
\end{equation}
where the amplitude is given by
\begin{eqnarray}
A(j_f,i_e)&=& 15j_{\scriptscriptstyle SO(4)}\!\!
\left(({\scriptstyle\frac{j_f}{ 2}},{\scriptstyle\frac{j_f}{2}}), f(i_e)
\right)\nonumber\\
&&\hspace{-4em} = \sum_{i_e^{\scriptscriptstyle +},
i_e^{\tiny -}} 15j_{\scriptscriptstyle SO(4)}\!\!
\left(({\scriptstyle\frac{j_f}{ 2}},{\scriptstyle\frac{j_f}{2}}),
i_e^{\scriptscriptstyle +},i_e^{\tiny -}\right)
\prod_{e\in v} f^{i_e}_{i_e^{\scriptscriptstyle +}
i_e^{\tiny -}}\ \ .
\label{A}
\end{eqnarray}\vskip-.1cm \noindent
Notation is as follows. The model is defined on a fixed 4d
triangulation $\Delta$. We do not discuss here the issue of the
recovery of triangulation independence (see \cite{libro,BC,freidel}). We
denote by $f,e,v$ respectively the faces, tetrahedra and 4-simplices of  $\Delta$. 
The choice of letters is motivated by the fact that
it is convenient to think in terms of the cellular complex dual to $\Delta$ 
(whose 2-skeleton defines the spinfoam):
triangles are dual to faces ($f$), tetrahedra to edges
($e$), and 4-simplices to vertices ($v$). The sum in
(\ref{Z}) is over an assignment of an \emph{integer} spin $j_f$
(that is, an irreducible representation of $SO(3)$) to each face
$f$, and over an assignment of an element $i_e$ of a basis in the
space of intertwiners to each edge $e$. We recall that an
intertwiner is an element of the $SO(3)$ invariant subspace of the
tensor product of the four Hilbert spaces carrying the four
representations associated to the four $f$'s adjacent to a given $e$. 
We use the usual basis given by the spin of the virtual link, 
under a fixed pairing of the four faces.  $\dim
j=2j+1$ is the dimension of the representation $j$.
$15j_{\scriptscriptstyle SO(4)}$ is the Wigner 15j symbol of the
group $SO(4)$. It is a function of 15 $SO(4)$  irreducible
representations.  A representation of $SO(4)$ can be written as a
pair of representations of $SU(2)$, in the form $(j^{\scriptscriptstyle +},j^{\tiny -})$, and
the $SO(4)$ 15j symbol is simply the product of two conventional
Wigner $SU(2)$ 15j symbols
\begin{equation}
15j_{\scriptscriptstyle SO(4)}(j_f^{\scriptscriptstyle +},
j_f^{\tiny -},i_e^{\scriptscriptstyle +},i_e^{\tiny -}) =
15j(j_f^{\scriptscriptstyle +},i_e^{\scriptscriptstyle +})\ 
15j(j_f^{\tiny -},i_e^{\tiny -}).
\end{equation}
The last object to define, and the key ingredient of our construction, 
is the linear map $f$ appearing in the
first line of (\ref{A}). This is a map from the space of the $SO(3)$
intertwiners between the representations $2j_1,...,2j_4$, to the
space of the $SO(4)$ intertwiners between the representations
$(j_1,j_1,),...,(j_4,j_4)$. The second line of (\ref{A}) simply
reexpresses this map in terms of its linear coefficients in the
basis chosen
\begin{equation}
f |i\rangle= \sum_{i^{\scriptscriptstyle +},i^{\tiny -}}  f^{i}_{i^{\scriptscriptstyle +}i^{\tiny -}} |  i^{\scriptscriptstyle +},i^{\tiny -} \rangle.
\label{f}
\end{equation}
These coefficients are defined as the evaluation
of the spin network   \begin{equation}
 f^{i}_{i^{\scriptscriptstyle +}i^{\tiny -}} = \begin{picture}(80,40)(-10,40)
   \includegraphics[height=3cm]{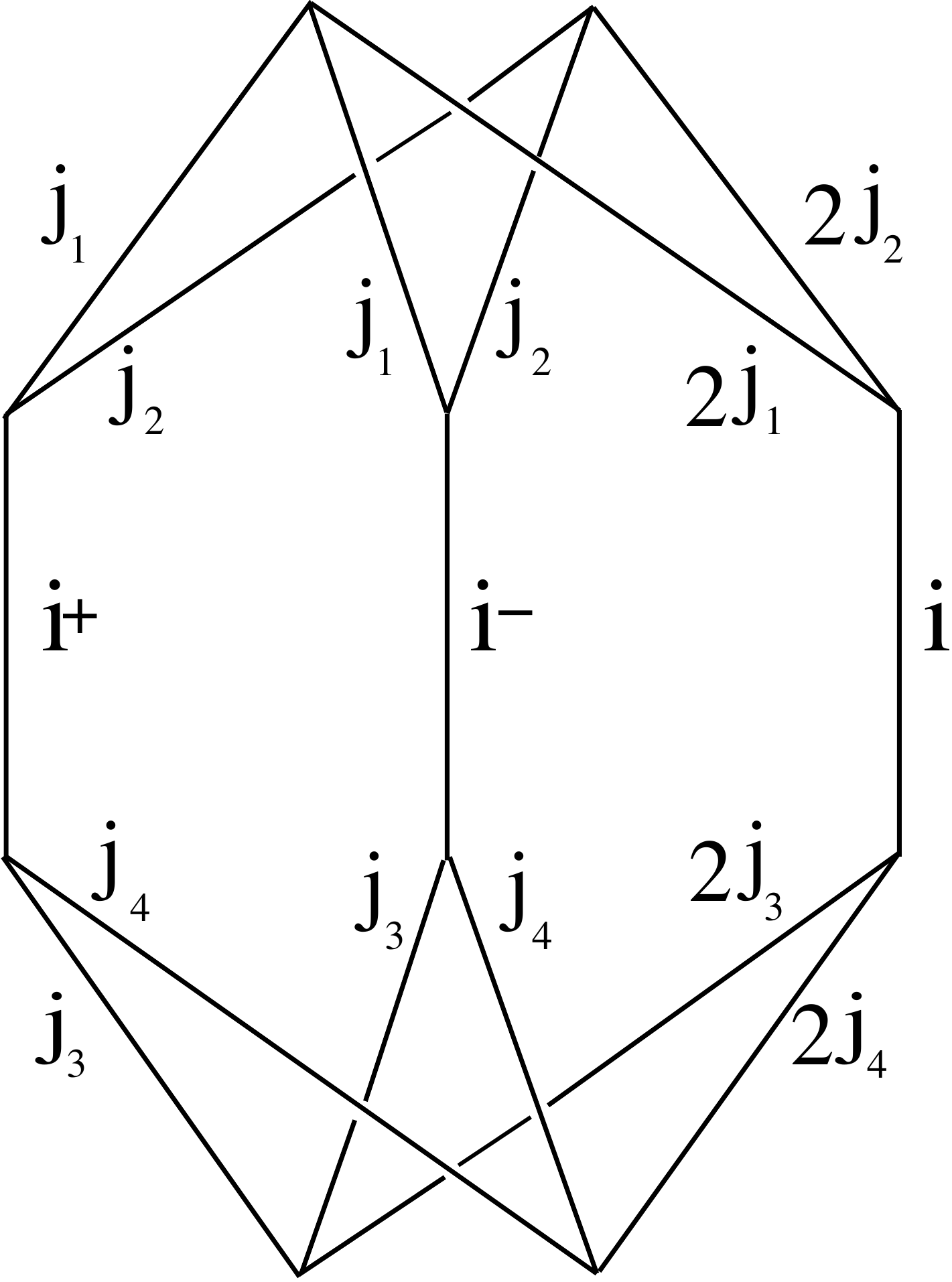}
\end{picture}
  \end{equation}\vskip1.5cm \noindent 
on the trivial connection. The amplitude can also be written in the form
\begin{equation}
A(j_f,i_e)\!=\!\!\int_{\!\!\!{}_{SU\!(2)^5}} \hspace{-1.5em}dV_e\ \big\langle\!\bigotimes_{ee'}\overset{\scriptscriptstyle  j\!\!{}_f\!\!{}_{/\!2}}{D}(V_e) 
\otimes \overset{\scriptscriptstyle  j\!\!{}_f\!\!{}_{/\!2}}{D}(V_{e'}^{-1})
,  \bigotimes_e i_e\!\big\rangle
\label{A2}
\end{equation}\vskip-.1cm \noindent
where index contraction is dictated by the standard 4-simplex graph
and the $j_f$ indices of the intertwiners are contracted with the $\frac{j_f}2\otimes \frac{j_f}2$
indices of the representation matrices $D$.
This concludes the definition of the model (for information on the general 
formalism, and more details on notation see \cite{libro}).  Let us now comment on its features.

First, the boundary states of the theory are spanned by trivalent graphs 
colored with $SO(3)$ spins and intertwiners.  Second, the model is a 
simple modification of the BC model as follows.  The BC model is given by
\begin{equation}
Z_{\rm BC}= \sum_{j_f}\ \prod_f (\dim j_f)^2\ \prod_v A_{BC}(j_f)
\label{Zbc}
\end{equation}
where here the sum is over half-integer spins and the amplitude is given by
\begin{eqnarray}
A_{BC}(j_f)&=& 15j_{\scriptscriptstyle SO(4)}\left((j_f,j_f),
i_{BC}\right).
\end{eqnarray}
The difference between the two theories is therefore in the intertwiner 
state space.  The relevant (unconstrained) intertwiner space is here the 
$SO(4)$ intertwiner space between four simple representations 
\begin{equation}
H_e={\rm Inv}(H_{(j_1,j_1)}\otimes...\otimes H_{(j_4,j_4)}).
\end{equation}  
The Barret-Crane intertwiner 
\begin{equation}
|i_{BC}\rangle =\sum_j (2j+1)|j,j\rangle
\end{equation}  
\vskip-.3cm
\noindent is a vector in this space.  The Barrett-Crane 
theory therefore constrains entirely the intertwiner degrees of freedom. 
In the model (\ref{Z}), instead, intertwiner degrees of freedom remain free. 
More precisely, the states (\ref{f}) span a subspace $K_e$ of $H_e$. 
The step from the single intertwiner $i_{BC}$ to the space $K_e$ 
is therefore the essential modification made with respect to the BC model.  
Why this step?

The reduction of the intertwiner space to the sole $i_{BC}$ vector
is commonly motivated by the imposition of the off-diagonal simplicity
constraints. For each couple of faces $f,f'$ adjacent to $e$,
consider the pseudoscalar $SO(4)$ Casimir operator
\begin{equation}
C_{ff'}=\epsilon_{IJKL}B^{IJ}_f B^{KL}_{f'}
\label{simpli}
\end{equation}
on the representation $(H_{(j_f,j_f)} \otimes H_{(j_{f'},j_{f'})})$. 
($\epsilon_{IJKL}$ is the fully antisymmetric object and
summation over repeated indices is understood.) Here $f\ne f'$ 
and $B^{IJ}_f$ with $I,J=1,...,4$ are the generators of $SO(4)$ 
in $H_{(j_{f},j_{f})}$.
In the context of the BC theory, these generators are the quantum
operators corresponding to the classical  
bivector associated to the face $f$. $C_{ff'}$
vanishes in the classical theory because the bivectors of the faces
a single tetrahedron span a 3d space and therefore their external 
products (\ref{simpli}) are clearly zero. These are the off-diagonal simplicity
constraints. (The diagonal simplicity constraint $C_{ff}=0$
constrains the representations associated to each $f$ to be simple.)
In BC theory, the constraints  $C_{ff'}=0$ are imposed strongly on
$H_e$, and the only solution of these constraint equations is
$i_{BC}$ \cite{mike}.   But these constraints
do not commute with one another, and are therefore second class.
Imposing second class constraints strongly is a well-known way of
erroneously killing physical degrees of freedom in a theory.  An
alternative way to rewrite the off-diagonal simplicity
constraints is the following.  As noted, these constraints impose the faces of
the tetrahedron to lie on a common 3d subspace of 4d spacetime.
Iff they are satisfied, there is a direction $n^I$
orthogonal to all the faces: the direction normal to the
tetrahedron. The $B_f$ have vanishing components in this
direction.   Choose coordinates in which $n^I=(0,0,0,1)$ and let
$i,j$ be indices that run over the first 3 coordinates only. Then we
have $2C_4\equiv B_f^{IJ}B_f^{IJ}= B_f^{ij}B_f^{ij}\equiv C_3$.  
The off-diagonal simplicity constraints can be written as the requirement that
there is a common direction $n$ such that $C=2C_4-C_3=0$ for
all the faces of the tetrahedron.   In the
quantum context, $C_4$ is the quadratic Casimir of $SO(4)$, with 
eigenvalues  $ j^{\scriptscriptstyle +}(j^{\scriptscriptstyle +}+1)\hbar^2+ j^{\tiny -}(j^{\tiny -}+1)\hbar^2$; while $C_3$  is the quadratic 
Casimir of the $SO(3)$ subgroup of $SO(4)$ that leaves $n^I$ 
invariant, with eigenvalues $ j(j+1)\hbar^2$, where we have momentarily 
restored $\hbar\ne 1$ units for clarity.
Can the constraint $C=2C_4-C_3=0$ be imposed  quantum mechanically  on $H_e$? 
A simple $SO(4)$ representation $(j,j)$ transforms under the $SO(3)$
subgroup in the representation $j\otimes j=0\oplus ... \oplus 2j$. 
Precisely in the $2j$ component, namely in the highest $SO(3)$
irreducible, this constraint (with suitable ordering: 
\begin{equation}
C=\sqrt{C_3+\frac{\hbar^2}{4}}-
\sqrt{2C_4+\hbar^2}+\frac{\hbar}{2}\ )
\end{equation}
is solved. Thus imposing the constraints on each face 
selects from $(H_{(j_{f_1},j_{f_1})}\otimes ... \otimes H_{(j_{f_4},j_{f_4})})$ 
the space formed by the tensor product of the highest $SO(3)$ irreducibles. 
So far this depends on \emph{which}  $SO(3)$ subgroup we have chosen; 
but if we project to the $SO(4)$ invariant-tensor space, then the dependence 
drops out because all $SO(3)$ subgroups in $SO(4)$ are conjugate to one another. 
In fact, what we obtain is precisely $K_e$.  Finally, it is easy to check that the 
off-diagonal  simplicity constraints are all weakly zero in this space: this follows
from the fact that they are antisymmetric in the $i^{\scriptscriptstyle +}, i^{\tiny -}$ indices, while 
the states (\ref{f}) are symmetric.

We close by sketching the derivation of this model as a
quantization of a discretization of GR (see \cite{Perez}). Fix an oriented 
triangulation and restrict the metric to be a Regge metric on this triangulation; 
that is, a metric which is flat within each 4-simplex, and where curvature is 
concentrated on the triangles. In order to describe this metric, we choose 
as variables a co-tetrad one-form $e^I(t)$ for each tetrahedron of the 
triangulation, and a co-tetrad one-form $e^I(v)$ for each simplex.
The two will be related by an $SO(4)$ group element $V_{vt}\equiv V_{tv}^{-1}$.
For each face in each tetrahedron, we define
$B_f(t)=\int_f \star (e(t)\wedge e(t))$, where the star is Hodge duality in
$R^4$.  $B_f(t)$ and $B_f(t')$ are related by $B_f(t)U_{tt'}=U_{tt'}B_f(t')$, where 
$U_{tt'}=V_{tv}V_{vt''}... V_{v^nt'}$ is the product of the group elements 
around the oriented link of $f$, from $t$ to $t'$. The bulk action can be written as
\begin{equation}
S_{bulk}[e]=\sum_f  Tr[B_f(t) U_f(t)]
\end{equation}\vskip-.2cm\noindent 
where $U_f(t)$ is the product of the group elements $V_{tv}V_{vt'}$ around 
the link of $f$. The boundary terms of the action can be written as
\begin{equation}
\label{bdryaction}
S_{boundary}[e]=\sum_f  Tr[B_f(t) U_{tt'}]
\end{equation}\vskip-.2cm\noindent 
where $U_{tt'}$ is the product of the group elements of the sole
part of the link which is in the triangulation.  We
take $B_f(t)$ and $V_{tv}$ as basic variables,
and take into account the constraints on $B_f$. 
These are the closure constraint
\begin{eqnarray}
\sum_{f\in t} B_f(t)=0
\label{con}
\end{eqnarray}
\vskip-.2cm\noindent
and the simplicity constraints (\ref{simpli}), for all $f, f'$ (possibly equal)
in $t$. (The constraints relating triangles that meet only at one point,
which appear in other formulations, are automatically solved by our
choice of variables.)

On the boundary of the triangulation, the boundary coordinates are the
$B_f(t)$ for the boundary triangles $f$. These have only two
adjacent tetrahedra $t,t'$ on the boundary.  The conjugate momentum 
(as can be seen from (\ref{bdryaction})) is a
group element for each $f$. Therefore the canonical boundary
variables are precisely the same as those of $SO(4)$ lattice gauge theory. We
can thus choose the Hilbert space of $SO(4)$ lattice gauge theory as
our unconstrained Hilbert space. This space can be represented as
the $L^2$ space on the product of one $SO(4)$ per triangle.  The two
$B_f$ variables at each $f$ are represented by the left and right
invariant vector fields on the group element at $f$, which are
related to one another in the same manner as the corresponding
classical quantities.  The closure constraint (\ref{con})
gives gauge-invariance at each tetrahedron, and
reduces the space of states to the space of the $SO(4)$ spin networks
on the graph dual to the boundary triangulation. The simplicity
constraints  (\ref{simpli}), as seen above, reduce each $SO(4)$ link
representation to a simple one, and the intertwiners spaces to
$K_e$. The resulting space of states is not only mathematical
isomorphic to the corresponding one of $SO(3)$ loop quantum gravity, but it
can also be physically identified with it, because we have an
explicit identification of the quantum operators on the two spaces
with the same classical analogues, such as the area of the faces.

Finally, coming to the dynamics, we can evaluate the amplitude of a
single 4-simplex $v$. Fixing the ten $B_{tt'}\equiv B_f(t)$ variables 
on the boundary, this can be formally written as
\begin{equation}
A[B_{tt'}]=\int dV_{vt} \,\, e^{i\sum Tr[B_{tt'} V_{tv}V_{vt'}]}.
\end{equation}
Transforming to the conjugate variables gives
\begin{eqnarray}
A[U_{tt'}] \hspace{-1mm}&=&\hspace{-2mm}\int\hspace{-1mm}dB_{tt'}
e^{-i\sum\!Tr[B_{tt'}
U_{tt'}]}\  A[B_{tt'}]\nonumber \\
&=&\hspace{-1mm}\int\hspace{-1mm}dV_{vt}  \prod_{tt'}
\delta(U_{tt'}V_{t'v}V_{vt}).
\end{eqnarray}
This is the amplitude. We can now transform back to the spin network basis, using the
$SO(4)$ spin network functions $\Psi_{j^\pm_{tt'},i^\pm_t}(U_{tt'})$\vskip-3mm
\begin{eqnarray}
A[j^\pm_{tt'},i^\pm_t] \!\!
&=&\!\!\! \int dU_{tt'}  \Psi_{j^\pm_{tt'},i^\pm_t}(U_{tt'})\ 
A[U_{tt'}] \nonumber\\
&=&\!\!\! \int  dV_{vt}\ \Psi_{j^\pm_{tt'},i^\pm_t}(V_{tv}V_{vt'}) 
\end{eqnarray}\vskip-.5mm\noindent
Performing the integral gives
\begin{eqnarray}
A[j^\pm_{tt'},i^\pm_t]
&=&15j_{\scriptscriptstyle SO(4)}(j^{\scriptscriptstyle +}_{tt'},j^{\tiny -}_{tt'},i^{\scriptscriptstyle +}_t,i^{\tiny -}_t).
\end{eqnarray}
Combining this $15j_{\scriptscriptstyle SO(4)}$ amplitude with the 
constraints discussed above, gives the model (\ref{Z})-(\ref{A}).

\centerline{---}

\noindent Thanks to  Alejandro Perez for help, criticisms and comments, and to Daniele Oriti for 
useful discussions. JE gratefully acknowledges support by an NSF International Research Fellowship under grant OISE-0601844.

\end{document}